# Enzyme kinetics: A note on negative reaction constants in Lineweaver-Burk plots

Sharmistha Dhatt# and Kamal Bhattacharyya*

Department of Chemistry, University of Calcutta, Kolkata 700 009, India

#pcsdhatt@gmail.com  *pchemkb@gmail.com

**Abstract:** Reaction constants in traditional Michaelis-Menten type enzyme kinetics are most often determined through a linear Lineweaver-Burk plot. While such a graphical plot is sometimes good to achieve the end, it is always better to go for a few numerical tests that can assess the quality of the data set being used and hence offer more reliable measures of the quantities sought, furnishing along with appropriate error estimates. In this context, we specifically highlight how cases may appear with negative reaction constants and explore the origin of such bizarre findings.

## 1. Introduction

An enzyme-catalyzed reaction is usually assumed to follow the Michaelis-Menten (MM) [1] mechanism, given by

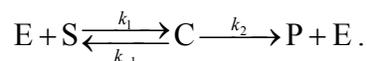
$$E + S \underset{k_{-1}}{\overset{k_1}{\rightleftarrows}} C \overset{k_2}{\longrightarrow} P + E.$$

Here, the substrate S combines with the enzyme E to form the complex C that subsequently yields the product P, regenerating the catalyst E. Denoting the concentrations of the species of interest by $e$, $s$, $c$, etc., the relevant rate equations are obtained as

$$dc/dt = k_1 e \cdot s - (k_{-1} + k_2)c; \quad dp/dt = V = k_2 c. \qquad (1)$$

Note that the rate $V$ in (1) and the concentrations of the species involved are actually time-dependent. In other words, $V = V(t)$, $c = c(t)$, etc. The initial conditions, $viz.$, $s = s_0$ and $e = e_0$ at $t = 0$, allow us to write the following two conservation equations, in addition

to (1), for completeness of the problem:

$$e_0 = e + c,$$
$$s_0 = s + c + p. \qquad (2)$$

All other rate equations follow from (1) and (2). From experiments, however, it is not easy to estimate the individual rate constants like $k_1$, $k_{-1}$ or $k_2$. Instead, one finds that two characteristic *reaction constants*, defined by

$$V_m = k_2 e_0, \quad K_m = (k_{-1} + k_2)/k_1, \qquad (3)$$

are obtainable. They turn out to be really important too in *all* discussions of enzyme kinetics [1-3]. In (3), $V_m$ refers to the maximum possible rate assuming, of course, that $e_0 \leq s_0$, while $K_m$ stands for the Michaelis constant.

The Lineweaver-Burk (LB) plot, loosely (see below) a plot of $V^{-1}$ vs. $s^{-1}$, offers, among others, a simple way of evaluating $V_m$ and $K_m$ from kinetic data. The plot is linear [though nonlinearity is apparent in (1)] and, by far, the most favorite one [1].

Over the years, therefore, considerable attention has been paid to problems like the accuracy of LB plot [2-5], the reasons behind deviations from linearity [6], the comparative performances with other linear plots [7], etc. In this respect, a few educational notes [8] are also available. Indeed, the LB scheme still remains a premier tool in studies on enzymatic reactions [9], and hence the endeavor to extract good reaction constants is continuing [10].

A survey of the relevant literature reveals, however, that discussions on the following points are still lacking: (i) How far does the nonlinear fitting procedure fare over the conventional LB numerical exercise? Is there any *quantitative* measure? (ii) Is there any *other* alternative to the nonlinear fitting? (iii) How should one ensure the *quality* of data set being handled? (iv) Why does one sometimes find *negative* reaction constants and when?

In what follows, we like to take up the above issues for clarifications. Obviously, our attention will be paid to numerical procedures, not to the drawing of the plots or their *visual* deviations from linearity.

## 2. The approximations

In the course of arriving at a suitable expression for the LB plot, a number of approximations are made. Of these, the first one is central to almost all discussions on the

MM kinetics. This is the steady state approximation (SSA). It assumes that

$$dc/dt \approx 0, \tau_1 \leq t \leq \tau_2. \tag{4}$$

The implication is, over a range of time, the complex concentration, and hence the rate, remains almost constant at $c = \bar{c}$. The SSA is invoked to simplify the expression for rate $V$. Let us note that, even if the SSA (4) does *not* hold, one can *still* find a time $t = \tau$ at some $c = \bar{c}$, corresponding to the maximum attainable value of $c$ [$\bar{c} \leq \min(s_0, e_0)$] for specific values of the rate constants and the initial conditions, that ensures

$$dc/dt = 0, t = \tau. \tag{5}$$

For convenience, the concentrations of other relevant species at this time ($t = \tau$) may be denoted by $\bar{e}, \bar{s}$, etc. Since (5) is more general than (4), we shall proceed with it. Putting this condition in (1), we obtain

$$k_1 \bar{e} \cdot \bar{s} - (k_{-1} + k_2)\bar{c} = 0; \quad \bar{V} = k_2 \bar{c}; \quad t = \tau. \tag{6}$$

Using the conservation for $e_0$ in (2), one finds from (6) the *exact* equation

$$\bar{V} = \frac{V_m \bar{s}}{K_m + \bar{s}}; t = \tau. \tag{7}$$

In favorable situations, (4) is obeyed; then, of course, (7) will *approximately* hold over the interval $\tau_1 \leq t \leq \tau_2$, but only during the *initial* times when $\bar{p} \to 0$. Thus, a more popular version of (7) reads as

$$\bar{V} = \frac{V_m s_0}{K_m + s_0}. \tag{7a}$$

It involves further approximations like $e_0 \ll s_0$ and $\bar{c}^2 \approx 0$. The significance of $\bar{V}$ in (7a) is just its constancy and *no reference to time* is explicit here. In essence, therefore, the key equation (7a) rests on the following assumptions: (i) The SSA is valid. (ii) The time $\tau_2$ is small enough to ensure that $\bar{p} \to 0$. (iii) The condition $e_0 \ll s_0$ is maintained. (iv) The stipulation $\bar{c}^2 \approx 0$ holds. (v) Finally, we need to also guarantee that the measured rate is *constant* within the concerned ($\tau_1 \leq t \leq \tau_2$) time interval.

For the sake of convenience, however, we simply write (7) or (7a) in the form

$$V = \frac{V_m s}{K_m + s}. \tag{8}$$

The reference to one of the two versions will be made when necessary.

**3. The schemes**

It is immediately clear that (8) can be rearranged to yield the LB equation

$$\frac{1}{V} = \frac{K_m}{V_m s} + \frac{1}{V_m}. \tag{9}$$

From (9), one notes that $1/V$ vs. $1/s$ will lead to a linear plot. It requires runs with different $s_0$ and such runs form a set. In the more precise case, we shall have to use $\overline{V}$ and $\overline{s}$ in places of bare $V$ and $s$ [as in (7)]. However, provided that the conditions listed below (7a) are obeyed, we may employ $s_0$ in (9) for $s$. Thus, we have *two sets* of data at hand. Calculations may now be categorized as follows:

***Scheme 1.*** Here, $1/V$ is fitted via a least squares method (LSM) as a function of $1/s$. The calculated slope and intercept then yield the reaction constants in (3). This is the scheme that is most popular. However, the plain graphical approach does not offer any quantitative estimate of error. Theoretically, an *overall measure* of error for the calculated $V_m$ and $K_m$ may be found from the expression

$$\varepsilon = \frac{1}{M} \sum_{j=1}^{M} \left| V_j - \left( \frac{V_m s_j}{K_m + s_j} \right) \right| \tag{10}$$

where $M$ observations form a set.

***Scheme 2.*** As the next option, one may go for a nonlinear fit (NLF) of data. In this scheme, one tries to minimize the error $\varepsilon$ in (10) by choosing suitable $V_m$ and $K_m$. It may be accomplished as follows. (i) Start from some arbitrary $K_m$. Calculate the two average values $N$ and $D$, defined by

$$N = \langle V \rangle, \; D = \langle s/(K_m + s) \rangle, \tag{11}$$

to obtain $V_m$ as $V_m = N/D$. (ii) Allow random variations of $V_m$ around this value to find a minimum value for $\varepsilon$. (iii) Note the values of $\varepsilon$, $V_m$ and $K_m$. (iv) Change $K_m$ systematically in both directions to repeat steps (i) to (iii) and record $\varepsilon$ only if it is less than the previous minimum. The recipe will finally yield a minimum $\varepsilon$ along with the *optimal* $V_m$ and $K_m$.

***Scheme 3.*** In view of the widespread belief that SSA is better obeyed with larger $s_0$, reducing, as a consequence, the corresponding $V^{-1}$ to be employed in LB calculations, it has been suggested that a better strategy would be to employ a *weighted* LSM (WLSM)

where $V^4$ acts as the weight factor [5]. Here, the graphical approach does not help. But, one may carry out the procedure in the same manner as the LSM. Only, the averages are computed with the weight factors. It is also recommended as a close alternative to the NLF [5]. Having obtained the values of $V_m$ and $K_m$ via this scheme, one may again stick to (10) for finding the overall error involved.

***Scheme 4.*** Keeping aside schemes like the LSM, WLSM or NLF, there is another possibility that does not seem to have been explored. In a set, we have $M$ points $(V_j, s_j)$. Consider, for example the case with $i = 1, j = 2$. Using (8), one can solve for the two unknowns from this pair to obtain the results

$$V_m(1,2) = \frac{s_2 - s_1}{s_2/V_2 - s_1/V_1}; \quad V_m(1,2)\left(\frac{1}{V_1} - \frac{1}{V_2}\right) = K_m(1,2)\left(\frac{1}{s_1} - \frac{1}{s_2}\right). \tag{12}$$

If the set is arranged in order of increasing $s$ and $V$, we notice from the second expression in (12) that $V_m(i,j)/K_m(i,j)$ is *always positive*. However, $V_m(i,j)$ is positive only when the restriction

$$s_j/s_i > V_j/V_i, \quad j > i, \tag{13}$$

is maintained. Otherwise, both $V_m$ and $K_m$ become *negative*. We shall see later the emergence of such a situation. For the points $(V_j, s_j)$, $j = 1, 2, .. M$, one will get $M(M-1)/2$ values of $V_m$ and $K_m$ on applying (12) in a pair wise manner. We call it the pair wise solution (PS) scheme. The results for $V_m$ and $K_m$ may be separately averaged, along with their standard deviations ($\delta V_m$ and $\delta K_m$). This PS scheme thus offers, unlike the three schemes sketched earlier, *independent* errors of the reaction constants.

**4. Sample results**

To explore the relative performances of the schemes, we need to choose the sets rather judiciously. In the present case, the rate constants are taken as $k_1 = 10.8$ μ mol$^{-1}$ min$^{-1}$, $k_{-1} = 10.0$ min$^{-1}$ and $k_2 = 0.8$ min$^{-1}$. We fix $e_0$ at 1 μ mol so that $V_m = 0.8$ μ mol min$^{-1}$ and $K_m = 1.0$ μ mol follow. These *exact* values will later be compared against the findings from different schemes. Henceforth, we shall continue with the corresponding units for other quantities, but may not explicitly refer to them, for brevity.

The numerical experiment is carried out via a fourth-order Runge-Kutta method applied to the coupled differential equations summarized in (1), with conservations (2).

The time step has been fixed at $10^{-6}$ min.

We start with $s_0 = 10$ μ mol and follow the dynamics up to 20 min. This is one run. Then, gradually, $s_0$ is increased in steps of 10 units. Six such runs are performed and the results are recorded. In Figure 1, we display the dynamics for a few runs. It is pretty clear that the SSA is *not always* obeyed. This is deliberate, but is required to show the intricacies of the LB scheme. Clearly, the SSA is worst for the case of $s_0$ equal to 10 units and is best when $s_0 = 60$.

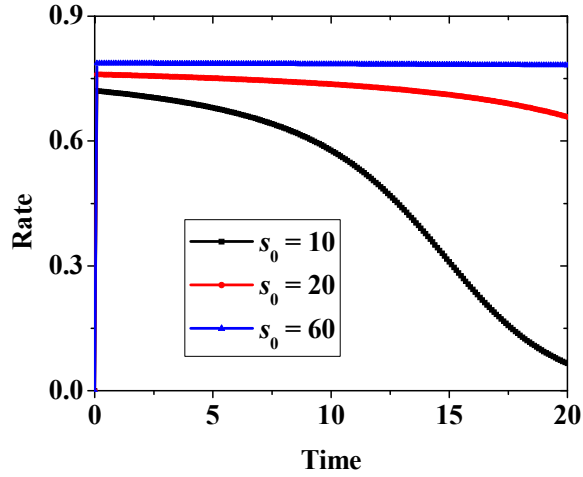

**Figure 1. Variation of $V(t)$ as a function of $t$ for three sample cases.**

In Table 1, we display the relevant results for our study. Here, the time $t = \tau$ refers

**Table 1. Results of runs at different times with varying initial substrate concentrations (in μ mol) $s_0$ at a fixed $e_0$ equal to 1 μ mol. The rate $V$ at a certain time $t$ and the substrate concentration at that instant $s$ are displayed, along with $s_0$ for the specific run.**

| $s_0$ | $t = \tau$ (1A) | | $t = 20$ m (2A) | |
|---|---|---|---|---|
| | $\overline{s}$ | $\overline{V}$ | $s$ | $V$ |
| 10.0 | 9.04 | 0.720 | 0.09 | 0.066 |
| 20.0 | 19.02 | 0.760 | 4.61 | 0.658 |
| 30.0 | 29.00 | 0.773 | 13.81 | 0.746 |
| 40.0 | 39.00 | 0.780 | 23.55 | 0.767 |
| 50.0 | 49.00 | 0.784 | 33.41 | 0.777 |
| 60.0 | 58.99 | 0.787 | 43.33 | 0.782 |

to the maximum point of the rate-vs.-time plot, *i.e.* where $(dV/dt) = 0$. Hence, (5) and

(7) are exactly obeyed. Thus, from the data for different runs summarized in Table 1, the following sets may be considered for study: (i) set of values for $((\bar{s}, \bar{V})$ that obeys (5) [to be called set 1A]; (ii) set of values for $((s_0, \bar{V})$ that is more traditional [to be called set 1B]; (iii) set of values for $((s, V)$ where $V(t)$ and $s(t)$ are measured at $t \gg \tau$ [to be called set 2A]; (iv) set of values for $((s_0, V)$ where $V(t)$ is *assumed* constant and measured at $t \gg \tau$ [to be called set 2B]. All these 4 sets are then employed to extract values of $V_m$ and $K_m$ by employing the schemes mentioned above. For consistency, we take $s$ and $V$ correct up to 2 and 3 decimal places, respectively, and report the reaction constants up to 3 decimal places. Results of Scheme 1 are displayed in Table 2.

Table 2. Results of employing the LSM to obtain $V_m$ and $K_m$. Sets A employ the actual values of $s$ at the times of rate ($V$) measurements, while sets B rely on the more traditional choice of $s_0$ in place of $s$. The sets are numbered in order of increasing time, displayed in Table 1.

| Set | $V_m$ | $K_m$ | $\varepsilon$ |
|---|---|---|---|
| 1A | 0.800 | 1.007 | 1.24 E-04 |
| 2A | 0.800 | 1.001 | 2.46 E-04 |
| 1B | 0.802 | 1.136 | 3.40 E-04 |
| 2B | -0.314 | -52.49 | 1.64 E-00 |

Let us note first that the performance is best for set 1A, as calculated $\varepsilon$ values reflect. Set 2B is not just the worst one; here, we encounter *negative reaction constants* as well. Note further that $V_m$ is usually found to be *more robust* than $K_m$. Set 2A performs nicely, much better than set 1B. Thus, the use $s(t)$ in (8) or (9) is preferred over the traditional use of $s_0$, the corresponding $V(t)$ at $t \gg \tau$ does not appear to cause very serious errors, provided we measure $s(t)$ and $V(t)$ for all runs *at the same t*. Considerable betterment, however, is achieved by adopting Scheme 2. These data are compiled in Table 3.

Table 3. Results of employing the NLF procedure on the sets.

| Set | $V_m$ | $K_m$ | $\varepsilon$ |
|---|---|---|---|
| 1A | 0.800 | 1.006 | 1.15 E-04 |
| 2A | 0.800 | 0.995 | 2.18 E-04 |
| 1B | 0.802 | 1.143 | 3.14 E-04 |
| 2B | 0.884 | 6.855 | 8.48 E-02 |

It is comforting to observe that the negative results have now disappeared from set 2B, though the error $\varepsilon$ is still considerable. Scheme 3 shows that it can also get rid of the negative reaction constants from 2B, and these results are displayed in Table 4. One may find additionally that the NLF fares only marginally than the WLSM, as had been rightly pointed out elsewhere [5].

Table 4. Results of employing the WLSM prescription on the sets.

| Set | $V_m$ | $K_m$ | $\varepsilon$ |
|---|---|---|---|
| 1A | 0.800 | 1.009 | 1.47 E-04 |
| 2A | 0.800 | 0.999 | 2.26 E-04 |
| 1B | 0.802 | 1.124 | 3.58 E-04 |
| 2B | 0.837 | 3.885 | 9.89 E-02 |

While we have seen so far that the relative performances of the LSM, NLF and WLSM can be judged *quantitatively* in terms of the error $\varepsilon$ given by (10), it is still not clear why *negative* reaction constants *appear* at all and how. To this end, we put forward Scheme 4 that acts as an *alternative* to the NLF. Data obtained via the PS for the two extreme (best and worst) sets 1A and 2B are shown respectively in Tables 5 and 6 below.

Table 5. Results of solving the equations pair wise on set 1A. The upper and lower entries display values of $V_m(i, j)$ and $K_m(i, j)$ respectively. Here, $i$ runs from 1 to 5 and $j > i$.

|   | 2 | 3 | 4 | 5 | 6 |
|---|---|---|---|---|---|
| 1 | 0.800<br>1.008 | 0.800<br>1.000 | 0.800<br>1.006 | 0.800<br>1.006 | 0.800<br>1.010 |
| 2 |   | 0.799<br>0.977 | 0.800<br>1.002 | 0.800<br>1.002 | 0.801<br>1.014 |
| 3 |   |   | 0.801<br>1.052 | 0.801<br>1.032 | 0.801<br>1.052 |
| 4 |   |   |   | 0.8<br>1.0 | 0.801<br>1.051 |
| 5 |   |   |   |   | 0.802<br>1.128 |

As expected, scatter in the estimated values of either $V_m(i, j)$ or $K_m(i, j)$ is little in case 1A. It is also important to witness that $V_m(i, j)$ values differ from the exact $V_m$ (equal to 0.8) only in the third decimal place, if at all, while the $K_m(i, j)$ data are mostly affected from true value at the second decimal place. Thus, we again notice the relative robustness of $V_m$ in comparison with $K_m$. Therefore, this seems to be a general feature [cf. Tables 2 –

5]; measured $K_m$ is more susceptible to error when we try to simplify (7) by choosing alternatives like (7a) or (8). This is exactly the situation with the set 2B. We exemplify in

**Table 6. Results of solving the equations pair wise on set 2B. The upper and lower entries display values of $V_m(i, j)$ and $K_m(i, j)$ respectively. Here, $i$ runs from 1 to 5 and $j = i + 1$.**

|   | 2 | 3 | 4 | 5 | 6 |
|---|---|---|---|---|---|
| 1 | -0.083<br>-22.510 | -0.180<br>-37.226 | -0.302<br>-55.746 | -0.459<br>-79.530 | -0.669<br>-111.295 |
| 2 |   | 1.018<br>10.954 | 0.919<br>7.942 | 0.884<br>6.855 | 0.863<br>6.242 |
| 3 |   |   | 0.838<br>3.690 | 0.829<br>3.324 | 0.822<br>3.042 |
| 4 |   |   |   | 0.820<br>2.751 | 0.814<br>2.442 |
| 5 |   |   |   |   | 0.808<br>1.995 |

Table 6 how dangerous such scatter may be for the worst set.

Origin of the *negative* reaction constants in case 2B should be transparent now. The failure to respect condition (13) is the real reason. We also notice that these cases follow only when the PS involves $i = 1$. This is not unexpected in view of the largest deviation from SSA shown in Figure 1 for the case of $s_0 = 10$, the first member of any set.

The PS scheme introduced here unveils not only the reason behind *negative* $V_m$ or $K_m$, but allows us to also obtain a kind of *average* estimate, along with the standard deviation. These results for the various sets are displayed in Table 7 below:

**Table 7. Average values of reaction constants and their standard deviations obtained from the PS scheme on the sets.**

| Set | $\langle V_m \rangle$ | $\delta V_m$ | $\langle K_m \rangle$ | $\delta K_m$ |
|---|---|---|---|---|
| 1A | 0.800 | 6.81 E-04 | 1.023 | 3.53 E-02 |
| 2A | 0.800 | 7.44 E-04 | 1.002 | 2.42 E-02 |
| 1B | 0.802 | 1.02 E-03 | 1.111 | 3.77 E-02 |
| 2B | 0.462 | 5.80 E-01 | -17.138 | 36.172 |

This table tells us about the *sanctity* of the estimated values. We should theoretically have $\delta V_m \ll |\langle V_m \rangle|$ and $\delta K_m \ll |\langle K_m \rangle|$. Note that, in this respect, set 2B is miserable, specifically due to its inability to respect the above condition for $K_m$. The situation is not

that bad for $V_m$, however. The averaging process has somehow managed to yield a positive value for it. Table 7 also furnishes *separate error estimate* for each of the two reaction constants. As an alternative, if we look for an overall error here too on a similar footing as the other cases considered earlier, we may go for the same by putting these average estimates in (10). In addition, we learn from such a table why set 2B should not be employed in calculating $K_m$ [and hence $V_m$]. In other words, *goodness of a set* may as well be judged on the basis of this PS table, in conjunction with Table 3.

## 5. Concluding remarks

In summary, we amplified via (10) how far the nonlinear fitting procedure fares over the conventional LB numerical exercise. This $\varepsilon$ provides a *quantitative* measure. We found also that WLSM is almost a virtual (poorer but simpler) alternative to the NLF. An altogether different kind of approach, the PS, has been advocated here. It shows very clearly why one sometimes finds *negative* reaction constants. An added advantage of this scheme is that here individual errors of $K_m$ and $V_m$ are found. On the basis of such observations, it is possible for us to finally ensure the *quality* of data set being handled.

## Acknowledgement

SD wishes to thank CSIR, India, for a research associateship.

## References

[1] K. A. Johnson and R. S. Goody, *Biochem.* 2011, 50, 8264.
[2] S. Schnell and P. K. Maini, *Comments Theor. Biol.* 2003, 8, 169.
[3] A. Cornish-Bowden, *Persp. Sci.* 2015, 4, 3.
[4] G. N. Wilkinson, *Biochem. J.* 1961, 80, 324; R. Eisenthal and A. Cornish-Bowden *Biochem. J.* 1974, 139, 715.
[5] A. Cornish-Bowden, *J. Theor. Biol.* 1991, 153, 437.
[6] P. C. Engel and W. Ferdinand, *Biochem. J.* 1973, 131, 97; L-H. Wang, M-S. Wang, X-A. Zeng, D-M. Gong and Y-B. Huang, *Biochim. Biophys. Acta*, 2017, 1861, 3189.
[7] J. E. Dowd and D. S. Riggs, *J. Biol. Chem.* 1965, 240, 863; G. L. Atkins and I. A. Nimmo, *Biochem. J.* 1975, 149, 775.
[8] N. C. Price, *Biochem. Edu.* 1985, 13, 81; R. J. Ritchie and T. Pravan, *Biochem. Edu.* 1996, 24, 196; J. K. Harper and E. C. Heider, *J. Chem. Educ.* 2017, 94, 610.
[9] W. Liang, A. P. Fernandes, A. Holmgren, X. Li and L. Zhong, *FEBS*, 2016, 283, 446.
[10] W. Stroberg and S. Schnell, *Biophys. Chem.* 2016, 219, 17; M. L. R. González, S. Cornell-Kennon, E. Schaefer and P. Kuzmič, *Anal. Biochem.* 2017, 518, 16.